\documentclass[aps,twocolumn,amsmath,amsfonts,nofootinbib,preprintnumbers,
superscriptaddress,eqsecnum,secnumarabic]{revtex4-1}
\usepackage[utf8]{inputenc}
\usepackage{color,graphicx,hyperref,amsmath,amssymb,multirow,slashed,ulem,mdwlist,soul,eufrak,threeparttable,diagbox}

\hypersetup{bookmarks=true,unicode=true,pdftoolbar=true,pdfmenubar=true,
	pdffitwindow=false,pdfstartview={FitH},pdftitle={VLQ},
	pdfauthor={G.~Cacciapaglia, T.~Flacke, M.~Park, M.~Zhang},
	pdfsubject={VLQ},
	pdfcreator={Creator},pdfproducer={Producer}, pdfkeywords={k1}{k2}{k3},
	pdfnewwindow=true,colorlinks=true,
	linkcolor=blue,citecolor=magenta,filecolor=magenta,urlcolor=cyan}

\begin{document}

%%%%%%%%%%%%%%% Begin Commands %%%%%%%%%%%%%%%%%%%%%%%%%%%%%%%%%%%%%%%%%%

\newcommand\sss{\scriptscriptstyle}
\newcommand{\be}{\begin{equation}}
\newcommand{\ee}{\end{equation}}
\newcommand{\bea}{\begin{eqnarray}}
\newcommand{\eea}{\end{eqnarray}}
\newcommand{\hc}{\text{h.c.}}
\newcommand{\Br}{\text{Br}}
\newcommand{\nn}{\nonumber}
\def\bsp#1\esp{\begin{split}#1\end{split}}

\newcommand{\red}[1]{{\color{red} #1}}
\newcommand{\blue}[1]{{\color{blue} #1}}
\newcommand{\tabincell}[2]{\begin{tabular}{@{}#1@{}}#2\end{tabular}}

	\title{Exotic decays of top partners: mind the search gap}
	
	\author{Giacomo Cacciapaglia}
	\affiliation{Institut de Physique des 2 Infinis (IP2I), CNRS/IN2P3 UMR5822, 69622 Villeurbanne, France}
	\affiliation{Univ Lyon, Universit\'e Claude Bernard Lyon 1, 69001 Lyon, France}
	\author{Thomas Flacke}
	\affiliation{Center for Theoretical Physics of the Universe, Institute for
		Basic Science (IBS), Daejeon 34126, Korea}
	\author{Myeonghun Park}
	\affiliation{Center for Theoretical Physics of the Universe, Institute for
		Basic Science (IBS), Daejeon 34126, Korea}
	\affiliation{Institute of Convergence Fundamental Studies and School of Liberal Arts, Seoultech, 232 Gongneungro, Nowon-gu, Seoul, 01811, Korea}
	\author{Mengchao Zhang}
	\affiliation{Center for Theoretical Physics of the Universe, Institute for
	Basic Science (IBS), Daejeon 34126, Korea}
	\affiliation{Department of Physics and Siyuan Laboratory, Jinan University, Guangzhou 510632, P.R. China}

	\date{\today}
	
	\begin{abstract}
	Many standard model extensions, including composite Goldstone Higgs models, predict vector-like fermionic top-partners at the TeV scale. The intensive search programmes by ATLAS and CMS focus on decays into a 3$^{\rm rd}$ generation quark and an electroweak boson ($W,Z,h$). However, underlying models of partial compositeness contain additional states that give rise to exotic top partner decays. We consider a well-motivated scenario in which a charge-$2/3$ top-partner decays into a pseudo-scalar, $T\rightarrow t\ a$, with $a\rightarrow gg \mbox{ or } b\bar{b}$ dominating below the $t\bar{t}$ threshold. We show that the constraints on the top partner mass from QCD pair production are substantially weakened, still allowing a top partner mass as light as $400$~GeV.
	\end{abstract}
	
	\maketitle

\maketitle

Top parters, i.e. vector-like quarks that couple to the top (and bottom) quarks, are a crucial ingredient in models of composite Goldstone Higgs~\cite{Kaplan:1983fs} with top partial compositeness~\cite{Kaplan:1991dc}. They are usually expected to have masses between $\approx 1$ to several TeV, with the lightest masses preferable if they play the role of regulators of the top loop corrections to the Higgs mass~\cite{Contino:2006qr}. Following the minimal coset~\cite{Agashe:2004rs}, they are expected to decay into a top or bottom, plus a Standard Model (SM) massive boson: for a top partner of charge $2/3$, $T$, the standard decay channels are thus $T\to b\ W^+$, $t\ Z$ and $t\ h$~\cite{DeSimone:2012fs}. Both ATLAS and CMS at the Large Hadron Collider (LHC) have extensive search programmes tailored to search for these states, leading to current bounds in the range $1.30 \div 1.42$~TeV depending on their branching ratios (BRs)~\cite{Aaboud:2018xuw, Aaboud:2018saj, Aaboud:2017qpr, Aaboud:2017zfn, Aaboud:2018wxv, Aaboud:2018uek, Aaboud:2018pii, Sirunyan:2018omb, Sirunyan:2017pks,Sirunyan:2019sza}.

Recently there has been raising interest in exploring ``exotic'' decays into non-SM bosons: they could arise as new pseudo-Nambu-Goldstones in non-minimal cosets~\cite{Serra:2015xfa,Bizot:2018tds,Han:2018hcu,Benbrik:2019zdp}, as additional pseudo-scalars carrying QCD colour charge~\cite{Bizot:2018tds}, as Dark Matter candidates~\cite{Anandakrishnan:2015yfa,Kraml:2016eti}, or they could be a gluon or a photon~\cite{Kim:2018mks,Alhazmi:2018whk} or simply new scalar states~\cite{Aguilar-Saavedra:2017giu,Chala:2017xgc,Dermisek:2019vkc,Benbrik:2019zdp}. Exotic decays of a charge $5/3$ top partner have been extensively studied in Ref.~\cite{Xie:2019gya}. While in this work we are interested in composite Higgs models, top partners also arise in other classes of models: extra dimensions (from which modern composite Goldstone Higgs models derive via the principle of holography~\cite{Contino:2003ve}~\footnote{Holographic interpretations of top partners can be found in Refs~\cite{Contino:2004vy,Cacciapaglia:2008bi}}), Little-Higgs models~\cite{Schmaltz:2002wx} (which can also be thought of as composite, see for instance~\cite{Low:2002ws}), and models where vector-like quarks are added ``by hand'' to the theory, like in supersymmetry~\cite{Aguilar-Saavedra:2017giu} and two-Higgs-doublet models~\cite{Chala:2017xgc,Dermisek:2019vkc,Benbrik:2019zdp}. Thus most of the results presented in this work can be extended to those scenarios.

\vspace{0.3cm}

In this letter we are interested in composite Goldstone Higgs models that feature an ``axion-like'' pseudo-scalar $a$ arising as a pseudo-Nambu-Goldstone boson from a global $U(1)$ symmetry. In models with an underlying gauge-fermion description~\cite{Ferretti:2013kya,Barnard:2013zea}, this state is ubiquitous~\cite{Ferretti:2016upr,Belyaev:2016ftv,DeGrand:2016pgq} and it can be potentially much lighter than the compositeness scale~\cite{Belyaev:2016ftv}.
In Ref.~\cite{Bizot:2018tds} it has been shown that it can have sizeable couplings to the top partners, in particular allowing decays of the charge-$2/3$ top partner $T \to t\ a$.
The Lagrangian of the pseudo-scalar $a$ is given by
\begin{eqnarray}
	{\mathcal{L}}_a &=& \frac{1}{2}(\partial_\mu a)(\partial^\mu a) 	-\frac{1}{2} m_a^2 a^2 + \!\sum_i\frac{g_i^2 K_i^a }{16\pi^2 f_a}a \mathcal{G}^i_{\mu\nu}\tilde{\mathcal{G}}^{i\mu\nu}\nonumber\\
    &&- \sum_f \frac{i C_f^a m_f}{f_a} a \bar f \gamma^5 f,
	\label{eq:aLag}
\end{eqnarray}
where $\mathcal{G}^i$ denote the $SU(3)_c\times SU(2)_L \times U(1)_Y$ field strengths~\footnote{Note that the $U(1)$ pseudo-scalar is the only state that can couple to both gluons and electroweak gauge bosons in gauge-fermion underlying models.} and $f$ denotes SM fermions. 
The decay constant $f_a$ is related to the composite Higgs decay constant $f_h$ that is expected to lie in the TeV range, and the  coupling constants $K^a_i$ and $C^a_f$ are dictated by the quantum numbers of the underlying fermions, and thus are fixed for each underlying model (we refer the reader to Ref.~\cite{Belyaev:2016ftv} for a comprehensive review and to Ref.~\cite{Cacciapaglia:2019bqz} for numerical tables of coefficients). 
While the precise numerical values vary among different models, the branching ratios share the following features: above the $t\bar{t}$ threshold, $m_a > 2 m_t$, the dominant decay is into tops, while below the dominant decay is in a pair of gluons, $a\to gg$, with decays into heavy fermions following, $a \to b\bar{b},\; \tau^+ \tau^-$.

The coupling to gluons allows for copious direct production of $a$ at the LHC, and bounds from direct searches have been exhaustively studied in Refs~\cite{Belyaev:2016ftv,Cacciapaglia:2017iws,Cacciapaglia:2019bqz}. Interestingly, there is a mass window $15~\mbox{GeV} \lesssim m_a \lesssim 65$~GeV where very weak constraints apply.~\footnote{Contrary to the generic analysis of Ref.~\cite{Bauer:2017ris}, decays of the Higgs boson $h\to aa$ and $h \to Za$ pose very weak constraints~\cite{Cacciapaglia:2017iws} in these models.} 
In the remaining mass range, significant bounds apply on $f_a$ from various channels, which, once converted to bounds on the Higgs decay constant $f_h$, are often stronger than those from electroweak precision tests.
For this reason we will focus in this letter on the low mass range and establish if significant bounds can derive from the production of $a$ via $T$ decays.
Are there gaps in the LHC search coverage for these signatures? How low can the bound on the $T$ mass be? Our main goal will be to answer these 	questions and guide the experimental effort toward a complete coverage of top partner signatures.

\section{Phenomenology of a top partner in presence of $T\rightarrow t\ a$ decays}

As a simplified model, we introduce the top partner $T$ with charge $2/3$, with the following Lagrangian~\cite{Bizot:2018tds}
\bea
\mathcal{L}_{T} &=& \phantom{+}  \overline{T}\left(i\slashed{D}-m_T\right) T + \left(\kappa^T_{W,L} \frac{g}{\sqrt{2}}  \,\overline{T}\slashed{W}^+P_Lb   \right. \nonumber \\ 
&&\left. + \kappa^T_{Z,L} \frac{g}{2 c_W}\, \overline{T} \slashed{Z} P_L t 
- \kappa^T_{h,L} \frac{m_T}{v}\, \overline{T} h P_L t   \right. \nonumber \\ 
&&\left.+ i \kappa^T_{a,L}\, \overline{T} a P_L t  + L\leftrightarrow R + \mbox{ h.c. }\right), \label{eq:LTa}
\eea
where $P_{L,R}$ are left- and right-handed projectors, and $T$ denotes the top partner mass eigenstate with mass $m_T$.  The first three interaction terms dictate the partial widths of $T$ decays into $b\ W$, $t\ Z$, and $t\ h$ as often considered in vector-like quark models~\cite{Buchkremer:2013bha,Barducci:2017xtw}, while the decay into $t\ a$ is the new ``exotic'' decay we consider. As shown in Ref.~\cite{Bizot:2018tds}, in underlying models with top partial compositeness the branching ratio $T\to t\ a$ can be sizeable and even dominate over the ``standard'' ones. Note that the couplings are always dominantly chiral, i.e. they either involve left-handed $t$ and $b$, or right-handed ones.

\begin{figure}[tbph]
	\begin{center}
		\includegraphics[width=0.5\textwidth]{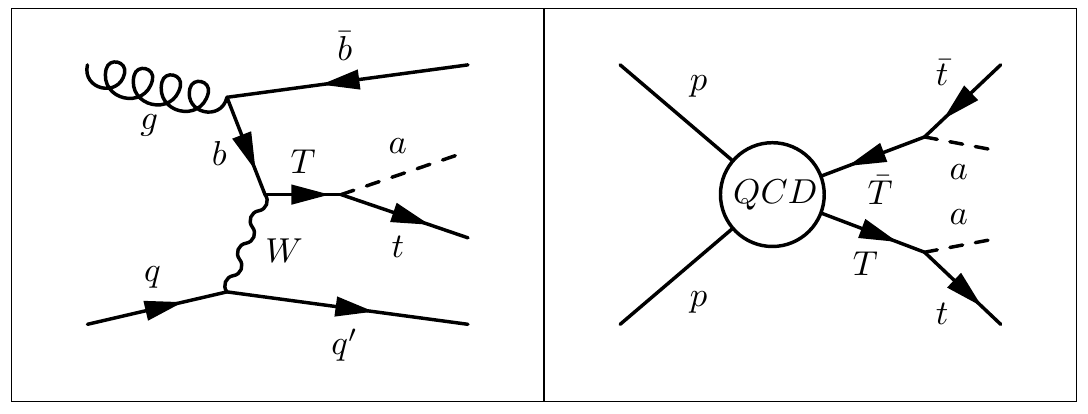}
		\caption{Example top partner production processes for electroweak single-production (left) and QCD pair production (right).}
		\label{fig:prod}
	\end{center}
\end{figure}

As the pseudo-scalar $a$ has small couplings to SM particles other than the top, it has a minor effect on $T$ production. The top partner $T$ is, therefore, single-produced in $b\ W$ and $t\ Z$ fusion (see Fig. \ref{fig:prod}, left) or pair-produced via its QCD interaction (see Fig. \ref{fig:prod}, right), as is commonly considered.  The presence of $a$ affects the searches by providing an additional decay channel, $T \rightarrow t\ a$, with the signatures depending on the decay modes of $a$. 
The cases of decays into $t\bar{t}$ and into a pair of electroweak gauge bosons have been studied in Refs~\cite{Han:2018hcu,Xie:2019gya,Benbrik:2019zdp}, showing a good coverage from current searches and prospects for improvements. In this letter we will focus on the more challenging decays $a\to gg$ and $a \to b\bar{b}$, which dominate at low mass. We should mention that searches targeting $\tau^+ \tau^-$ may also be relevant~\cite{Cacciapaglia:2017iws} as well as the more suppressed di-muon channel~\cite{Aaij:2017rft,Sirunyan:2018wim}.

We will focus on pair production due to QCD interactions, which has the benefit of only depending on the mass $m_T$.
To a good approximation, pair production with subsequent decays into $t\ a$, $t\ h$, $t\ Z$, $b\ W$ is characterised by three branching ratios (assuming that there are no additional channels), while the kinematics of the new channel depend on $m_a$ and the dominant decay channel of $a$.
We will assume here that $T$ has a small width so that production and decay can be factorised\footnote{See Ref.~\cite{Moretti:2016gkr} for large-width effects in vector-like quark pair production.}, and under this approximation it is justifiable to neglect single production which is proportional to the couplings $\kappa^{T}_{W/Z/h/a,L/R}$ in eq.~\eqref{eq:LTa}. Moreover, we will neglect the effect of the top and bottom polarisation, which only affects the final state kinematics in a minor way.

ATLAS and CMS pursue an active search program for vector-like quarks, with several searches for pair-produced $T$ targeting decays into $T\rightarrow t\ h,\, t\ Z,\, b\ W$ extended to the $\sqrt{s}=13$~TeV data~\cite{Aaboud:2018xuw, Aaboud:2018saj, Aaboud:2017qpr, Aaboud:2017zfn, Aaboud:2018wxv, Aaboud:2018uek,  Sirunyan:2018omb, Sirunyan:2017pks,Sirunyan:2019sza}. Both collaborations provide bounds in benchmarks as well as  in ``branching ratio triangle'' summary plots, assuming $\mbox{BR}(T\rightarrow t\ h)+ \mbox{BR}(T\rightarrow t\ Z)+\mbox{BR}(T\rightarrow b\ W)=1$. ATLAS provides a combination of its individual searches~\cite{Aaboud:2018pii} which established a lower bound on $m_T$ of $1300 \div 1420$~GeV, depending on the $T$ branching ratios, with the strongest bound applying for $\mbox{BR}_{T\rightarrow th}=100\%$. The CMS bounds are comparable, with $m_T > 1370 \, / \, 1300 \, / \,  1295$~GeV for 100\% BR into $th$ / $tZ$ / $bW$ \cite{Sirunyan:2018omb, Sirunyan:2017pks,Sirunyan:2019sza} \footnote{The CMS searches are not combined, yet.}.

How are these bounds modified in the presence of the $T\rightarrow t\ a$ decay?  Pair production searches are still mainly based on a cut-and-count method. The number of signal events in a given signal region is
\begin{equation} \label{eq:Ns}
N_{\rm{s}}=\mathcal{L} \, \sigma_{T,\rm{p}}(m_T) \sum_{ij}\epsilon_{ij}(m_T) \, \mbox{BR}_i \, \mbox{BR}_j\,,
\end{equation}
where $\mathcal{L}$ is the integrated luminosity used in the search, $\sigma_{T,\rm{p}}$ is the $pp\to T\bar{T}$ production cross section (which depends on $m_T$), and $i,j = W, Z, h, a$ label the different decay channels of $T$.  The factor $\epsilon_{ij}$ is the signal efficiency for a pair-produced $T$ with one $T$ decaying through decay $i$ with branching $\mbox{BR}_i$ and its charge conjugate through decay $j$ with branching $\mbox{BR}_j$.
Fully determining the bound on $m_T$ would require knowledge of all $\epsilon_{ij}(m_T)$ for each search signal region. Typically, the searches are designed in such a way that each signal region targeting a specific final state, $t\ Z$, $t\ h$ or $b\ W$, has very low sensitivity to the others, so that the signatures are picked up exclusively by one search. However, this is not the case for the new channel, which can in principle populate different signal regions. One conservative approach consists in ignoring the new channel and rescale the bound with the reduced BR in the standard channels~\cite{Aguilar-Saavedra:2017giu}, and a discussion is currently undergoing within the experimental collaborations on ways to update the triangle summary plots\footnote{Private communication.}. In this letter we will discuss how the new signature affects the bound in some motivated benchmark points by recasting the relevant searches.

\section{Bounds for exclusive decays, $\mbox{BR} (T\rightarrow t\ a) = 100\%$, $a\to gg,\; b\bar{b}$}

We first focus on exclusive decays in the new channel, and we identify 3 searches that are sensitive to decays $a\to gg$ and $a \to b\bar{b}$. As our main interest is on the low mass region, the final states may be boosted thus leading to merging jets.
The three searches we recast are:

\begin{itemize}
	\item For very light $a$, the two jets are strongly collimated and $a$ can be mis-identified as a single QCD jet. The final state is, therefore, similar to $t\bar{t}jj$ and we recast the CMS search for ``excited tops'' at $\sqrt{s}=13$~TeV ~\cite{Sirunyan:2017yta} to cover this region. 

	\item For larger $m_a$, the di-jet system becomes resolved and, for hadronically decaying tops, the final state contains many energetic jets. Searches for R-parity violating supersymmetry (RPV SUSY) in hadronic final states can cover the target signature. 
We thus recast the ATLAS $\sqrt{s}=8$~TeV search from Ref.~\cite{Aad:2015lea}. The most recent $\sqrt{s}=13$~TeV in Ref.~\cite{Aaboud:2018lpl} applies a much stronger cut on the summed jet-mass, which rejects most of the signal and, therefore, is less sensitive to our final state (see Appendix \ref{app:RPV13} for details). Similarly, the CMS $\sqrt{s}=13$~TeV RPV SUSY search \cite{Sirunyan:2017dhe} requires a high summed jet-mass and a high $H_T$ cut.

	\item If $m_a$ is close to the $W$-$Z$ or Higgs masses, it can be mis-tagged as a hadronically decaying SM boson. Thus, standard top partner searches could have some residual sensitivity to the new channel. We found that the ATLAS search for $T \to t\ h$ in Ref.~\cite{Aaboud:2018xuw} is recastable and most sensitive to the signal for $a \to b \bar{b}$: it is based on a broad-band search with multiple signal regions using 0 or 1 lepton, many $b$ tags, and cut-based hadronic top- and/or $h_{bb}$ tags, which are loose enough to capture some of the new signal. Top partner searches tagging hadronic $W$ and $Z$~\cite{Aaboud:2017zfn,Aaboud:2018uek,Aaboud:2018wxv} use a 50\%-efficient W-tagging working point~\cite{ATL-PHYS-PUB-2015-033} or a deep neutral network to tag boosted objects. They are potentially sensitive to $a\to gg$, however determining the tagging efficiency for $a$ is very difficult with the available information. 
	
\end{itemize}
Details and validation of the recasts are summarised in the Appendices.

\begin{figure*}[htbp]
	\begin{center}
		\includegraphics[width=0.90\textwidth]{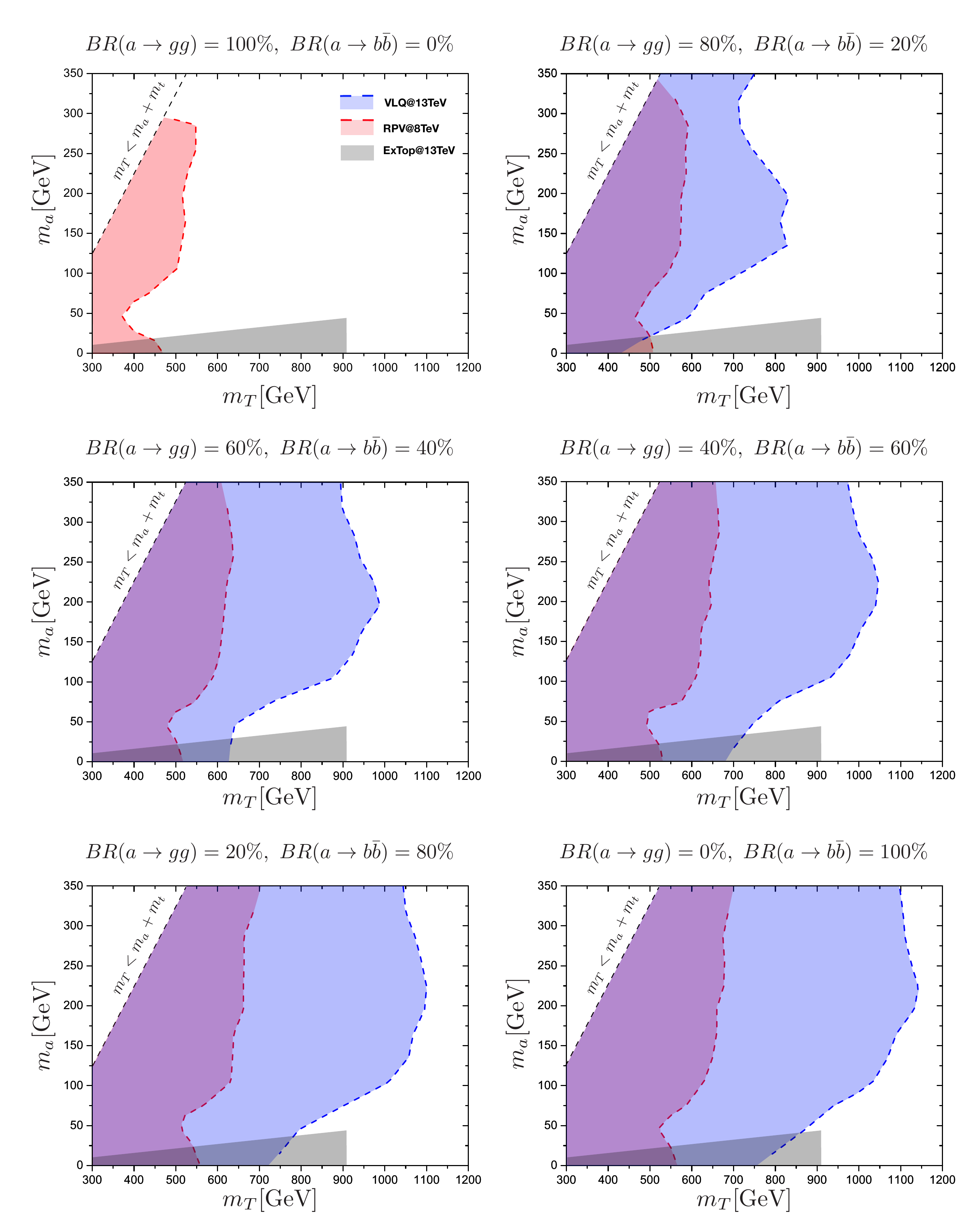}
		\caption{Direct search limits on $m_T$ vs $m_a$ plane for different branching ratios of $a$ to $gg$ and $b\bar{b}$. In the  region $m_T < m_a+ m_t$, the $T\rightarrow t a$ decay is kinematically forbidden. The grey region at low $m_a$ is excluded by CMS excited top search \cite{Sirunyan:2017yta}, as the boosted light $a$ is tagged as a single jet. The red region is excluded by 8TeV ATLAS multi-jets RPV SUSY search \cite{Aad:2015lea}. The blue region is excluded by 13~TeV ATLAS VLQ search \cite{Aaboud:2018xuw}. Due to  b-jet tagging requirements in the signal regions, limits to the model with a higher $BR(a\to b \bar{b})$ are more stringent. }
		\label{Recast}
	\end{center}
\end{figure*}

In Fig. \ref{Recast} we present the 95\% C.L. limits in the $m_T$ vs $m_a$ plane for the three recast searches and for various combinations of BRs $a\rightarrow gg$ and $a\rightarrow b \bar{b}$. The grey areas are excluded by the excited top search~\cite{Sirunyan:2017yta}, where $a$ is tagged as a single QCD jet. They only cover very low masses, well below $50$~GeV, with a slight dependence on the $T$ mass, and the bound extends up to $m_T \approx 900$~GeV.
The red regions in Fig. \ref{Recast} are excluded by the 8~TeV RPV SUSY search~\cite{Aad:2015lea}. 
The most sensitive signal region requires 7 jets with $p_\text{T}$ larger than $80$~GeV and at least two $b$-tagged jets, therefore the bound is the strongest for $\mbox{BR}(a \to bb) = 100\%$. Our bounds for $\mbox{BR}(a \to gg) = 100\%$ are significantly weaker than those in Ref.~\cite{Anandakrishnan:2015yfa} for the same signal: while we recast the same search, we used the published results that have a larger estimate of the background compared to the note used by Ref.~\cite{Anandakrishnan:2015yfa}.~\footnote{We checked that our recast is compatible with that of Ref.~\cite{Anandakrishnan:2015yfa}.} 
Finally, the blue regions in Fig. \ref{Recast} are excluded by $13$~TeV $T \to t\ h$ search~\cite{Aaboud:2018xuw}.  
We recast the ``1-lepton regions'', which target $T \to t\ h_{bb}$ by demanding at least 3 $b$-jets.~\footnote{The other ``0-lepton'' regions require large missing energy from invisible decays of the $Z$, thus they are not sensitive to the new channel.}
This search is, therefore, insensitive to the $a\to gg$ channel, however it gives  strong bounds for $a\to b \bar{b}$ decays. It can be seen in Fig. \ref{Recast} that the reach is maximised for $m_a \approx m_h$, for which the search is optimised. Yet, this $T$ search can dominate the bound even for sub-dominant $b\bar{b}$ BR. We should note that our recast underestimates the bound on the $T\to t\ h$ channel mainly due to a marginally smaller signal efficiency and the fact that we do not combine different signal regions (see Appendix~\ref{app:VLQ} for more details).

%%%%%%%%%%%%%%%%%%%%%%%%%%%%%%%%%%%%%%%%%%%%%%%%%%%%%%%%%%%%%%%%%%%
%%%%%%%%%%%%%%%%%%%%%%%%%%%%%% FIGURE  %%%%%%%%%%%%%%%%%%%%%%%%%%%%%%%%
%%%%%%%%%%%%%%%%%%%%%%%%%%%%%%%%%%%%%%%%%%%%%%%%%%%%%%%%%%%%%%%%%%%
\begin{figure*}[bhtp]
	\begin{center}
		\includegraphics[width=0.90\textwidth]{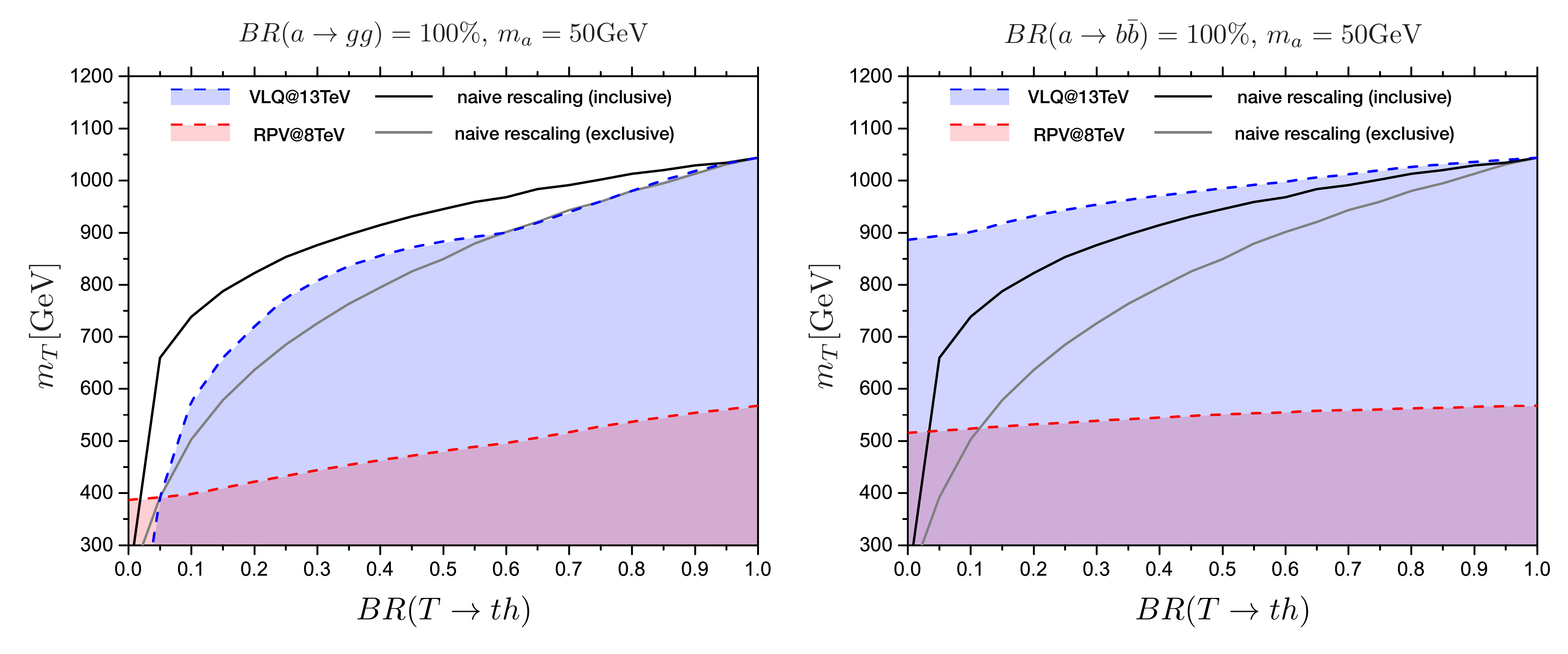}
		\caption{Direct search limits on $m_T$ as a function of $\mbox{BR}(T\to th)$. The red region is excluded by the $\sqrt{s} =8$~TeV ATLAS multi-jets RPV SUSY search~\cite{Aad:2015lea}, while the blue region is excluded by the $\sqrt{s}=13$~TeV ATLAS top partner search~\cite{Aaboud:2018xuw}. In both plots we fix $m_a=50$~GeV. 
		We also show bounds from naive rescaling. Black solid line and gray solid line correspond to inclusive search and exclusive search respectively. }
		\label{Limit}
	\end{center}
\end{figure*}
%%%%%%%%%%%%%%%%%%%%%%%%%%%%%%%%%%%%%%%%%%%%%%%%%%%%%%%%%%%%%%%%%%%
%%%%%%%%%%%%%%%%%%%%%%%%%%%%%% FIGURE  %%%%%%%%%%%%%%%%%%%%%%%%%%%%%%%%
%%%%%%%%%%%%%%%%%%%%%%%%%%%%%%%%%%%%%%%%%%%%%%%%%%%%%%%%%%%%%%%%%%%

Our results clearly show that for dominant $a\to gg$ decays, which is the norm in realistic models~\cite{Belyaev:2016ftv}, the bound on $m_T$ can be very weak: for $m_a\simeq 50$~GeV (which is not excluded by any direct searches~\cite{Cacciapaglia:2019bqz}), $m_T\lesssim 400$~GeV is still allowed while for larger $m_a$ the bound generically never passes $\approx 550$~GeV. This represents a gap in the top partner coverage, which could be closed by dedicated searches tagging low-mass di-jet resonances. Note that this final state resembles a target signal with hadronic $Z$ bosons, $tZ_{\rm{had}}\bar{t}Z_{\rm{had}}$. Another possible improvement would be to reduce the cuts on jet activity in the $\sqrt{s} = 13$~TeV search~\cite{Aaboud:2018lpl} to be able to cover the low mass region.

For dominant $T\rightarrow t\ a\rightarrow t\ b\bar{b}$, the bound for $m_T$ exceeds $1$~TeV for sufficiently heavy $a$. For light $a$ (which implies collimated $b\bar{b}$ pairs), the $T$ search looses sensitivity as search regions either demand a Higgs-tagged $b\bar{b}$-pair, or at least 4 sufficiently isolated $b$-jets. The coverage of this final state could also be improved by training an $a$-tagger algorithm in the low mass region.

\section{Bounds for $\mbox{BR} (T\rightarrow t\ a) + \mbox{BR} (T\rightarrow t\ h) = 100\%$}

The most general case where the pair-produced $T$ can decay in all combinations of the four channels $b\ W$, $t\ Z$, $t\ h$ and $t\ a$ may seem daunting to analyse. However, the standard top partner searches are designed to cover final states when either one or both $T$ decay in the focus channel, for instance $b\ W+X$ or $t\ Z+\bar{t}\ Z$, without contamination. If the signal regions have no sensitivity to the new channel $t\ a$, as is the case for leptonic $Z$ or $W$ decays for instance, then the presence of the new channel can be seen as an overall reduction of the effective cross section.
For an ``exclusive'' search region which is sensitive to a particular combination of $T\bar{T}$ decays, the number of signal events is reduced by a factor of $(1-\rm{BR}_a)^2$. For an ``inclusive'' search region which targets final states $t\ Z+X$, $t\ h+X$ or $b\ W+X$ (with a specific decay of one top partner whilst not having strong requirements on the second), the number of events is reduced by a factor
\begin{equation} \label{eq:naive}
\frac{\left. \sigma_{T,p} \right|^{\rm eff}}{\sigma_{T,p}} = (1-\mbox{BR}_a) + \frac{\epsilon_{\rm{td}}}{2} \mbox{BR}_a \mbox{BR}_{\rm{td}} + \mathcal{O}(\epsilon^2_{\rm{td}})\,,
\end{equation}
where $\mbox{BR}_{\rm{td}}$ is the branching ratio of $T$ into the targeted decay channel in presence of $\mbox{BR}_a$, and $\epsilon_{\rm{td}}$ is the signal efficiency to detect its decay products. These formulae  allow for a simple rescaling of the bound.

As mentioned in the previous section, this simple estimate can fail in the case of $T\to t\ h$, as searches based on $h \to b\bar{b}$ can also be sensitive to $a\to b\bar{b}$, even for different masses as shown in Fig.~\ref{Recast}.
In this section, we  therefore study the interplay of $t\ h$ and $t\ a$ decays in more detail.  
In particular,  we consider a benchmark scenario in which $T$ has two decay channels: $T\to t\ a$ and $T\to t\ h$. We fix $m_a = 50$~GeV, which corresponds to the weakest constrained mass value.
For the decay of $a$, we consider two limiting scenarios: $\mbox{BR} (a\to gg) = 100\%$ and $\mbox{BR}(a\to b\bar{b}) = 100\%$. 
The results of the scan are shown in Fig.~\ref{Limit}, where we demonstrate how the limit on $m_T$ evolves for increasing $\mbox{BR} (T\to t\ h)$. 
In the case of $a\to gg$, low masses below $800$~GeV are still allowed as long as $\mbox{BR}_h \lesssim 20\%$, while the mass bound remains strong otherwise and for $a\to b\bar{b}$. As a reference, we also show as a solid lines the naive estimate for the bound obtained by rescaling the number of signal events whilst assuming the search regions to be inclusive (top line, black) or exclusive (bottom line, grey). The reduction is translated into a bound on $m_T$ under the assumption that the signal efficiency is approximately constant in the relevant $m_T$ regime.
As can be seen in Fig.~\ref{Limit} (left), the decrease of the bound with decrease of $\mbox{BR}(T\rightarrow t\ h) = 1- \mbox{BR}(T\rightarrow t\ a)$ follows the shape expected for an exclusive search above $\mbox{BR}(T\rightarrow t\ h) \sim 0.6$ while it follows the shape expected for an inclusive search region below, indicating that the most constraining search regions differ in these two parameter regimes, and that the search has no explicit sensitivity to $T\rightarrow t\ a \rightarrow t\ gg$.  Fig.~\ref{Limit} (right) clearly deviates from the naive scaling relation as the decay $T\rightarrow t\ a \rightarrow t\ b\bar{b}$ populates the signal regions.

\section{Conclusions and outlook}

Current LHC searches for vector-like top partners constrain a charge $2/3$ top partner $T$ to be heavier than $1.3 \div 1.4$~TeV, depending on the pattern of decays to the standard channels $t\ Z$, $t\ h$, or $b\ W$, exclusively.
However, underlying models featuring partially composite tops always contain additional -- potentially light -- scalars, which provide common new decay modes for the top partners. In particular, all models with an underlying confining gauge group and underlying fermions predict the existence of a light pseudo-Nambu-Goldstone boson, SM singlet, $a$ that, if lighter than $2 m_t$, decays dominantly to $gg$ or $b\bar{b}$. This provides a theoretically well-justified motivation to focus on the exotic decays $T\rightarrow t\  a \rightarrow t\ gg$ and $T\rightarrow t\ a \rightarrow t\ b\bar{b}$.

In this letter, we studied top partner pair production via QCD interactions, followed by the exotic decays. We surveyed $T$ pair production searches as well as other searches by ATLAS and CMS, and identified three searches that can be sensitive to the new final states: a broad-band $T\to t\ h$ search, a $\sqrt{s}=8$~TeV RPV SUSY search, and the excited top-pair search. Besides being the most promising existing searches, they are easily recastable. The bounds we obtained show that the $T\rightarrow t\ a\rightarrow t\ gg$ channel is very weakly constrained, allowing for $m_T > 550$~GeV for any $m_a< 2 m_t$ and going as low as $m_T \approx 390$~GeV for $m_a=50$~GeV. The channel $T\rightarrow t\ a\rightarrow t\ b\bar{b}$ is better constrained due to the sensitivity of the broad-band $T\to t\ h$ search, although for light $m_a\lesssim 75$~GeV it looses sensitivity, allowing for $m_T$ down to $\approx 900$~GeV. We also provide estimates on the reduction of bounds if standard and exotic decays are both present, and study in detail the case of coexisting $T\rightarrow t\ h$ and $T\rightarrow t\ a$ decays in more detail.
The latter is important as the search for $T\to t\ h$ also covers the exotic channel $T\to t\ a \to t\ b\bar{b}$, so a combined analysis is necessary.

Our analysis shows that, although exotic top partner decays can exhibit spectacular and well-constrained final states, there exist well-motivated channels that are very hard to constrain with current searches. In the case $T\to t\ a \to t\  g g$ we found that the bound on the $T$ mass is strongly weakened, allowing values down to  $\approx 400$~GeV.
Nevertheless, it is possible to increase the sensitivity to these final states by accordingly modifying the current searches. For example, the RPV SUSY search at $\sqrt{s}=13$~TeV became insensitive to the $a\to gg$ channel because of a too-high cut on hadronic activity, which could be lowered in the future. Furthermore, $a\to jj$ decays resemble the hadronic decays of the $W$/$Z$ and the Higgs in the SM, thus the sensitivity could also be improved by hadronic taggers, either cut-based  or via BDT/machine learning techniques, which could identify di-jet resonances (boosted or resolved) with different masses. This problem is most severe for light $a$, with $m_a \lesssim m_W$, while heavier masses are covered, as shown by the reach of the $T\to t\ h \to t\ b\bar{b}$ broad-band search we recast. In the case of $T\to t\ a$, it is the over-specialised/trained cuts that ``{\it pour the baby (signal) out with the bath water}''. Future well-documented (recastable) broad-band searches or dedicated search regions targeting decay chains with $gg$ / $b\bar{b}$ resonances with $m_{jj}\neq m_W, m_Z$ and/or $m_{bb}\neq m_h$ could close this gap.

\acknowledgements
We thank Maxim Perelstein for feedback and clarifications on Ref.~\cite{Anandakrishnan:2015yfa} and Ref.~\cite{Aad:2015lea}. 
GC also thanks G.Ferretti, D.Buarque Franzosi, V.Ellajosyula and L.Panizzi for useful discussion.
GC acknowledges support from the France-Korea Particle Physics Lab (FKPPL) under project code CompHS, the Labex-LIO (Lyon Institute of Origins) 
under grant ANR-10-LABX-66 (Agence Nationale pour la Recherche), and FRAMA (FR3127, F\'ed\'eration de Recherche ``Andr\'e Marie Amp\`ere'').
TF, MP, and MZ were supported by IBS under the project code, IBS-R018-D1.

\vspace{1cm}

\newpage

\appendix

\section{Validation of recasts}\label{app:recast}

All events are generated with {\tt MadGraph5}~\cite{Alwall:2014hca}, then showered and hadronised with {\tt PYTHIA 8}~\cite{Sjostrand:2014zea}. 
{\tt DELPHES 3} is used for a fast detector simulation~\cite{deFavereau:2013fsa}, and {\tt FastJet} is used for jet clustering~\cite{Cacciari:2011ma}.
The $T\bar{T}$ production cross-section at $\sqrt{s} =8$~TeV and $13$~TeV LHC are calculated by {\tt Top$++$}~\cite{Czakon:2011xx}. 
In the following, we provide details of the recast and validation of the cut flows of the searches used in this article.

\subsection{CMS 13~TeV excited top quark search}

When the pseudo-scalar $a$ is light and highly boosted, it resembles a single QCD jet. Thus, the effective final state of $T$ pair production with a $T\rightarrow ta \rightarrow tgg$ or $tb\bar{b}$ decay is $t\bar{t}jj$, which is targeted by the search for excited top quark pair production~\cite{Sirunyan:2017yta}.  This search, however, is not cut-flow based and a recast is not possible: we will therefore assume that the $a$ decays are tagged as a single jet as long as the separation angle between the two jets is small enough.

To provide an estimate on the mass limit below which the pseudo-scalar $a$ can be treated as single jet,
we calculated the momentum $a$ obtains from a $T$ decaying at rest:
\begin{eqnarray}
\nonumber p &=& \frac{1}{2m_T} \sqrt{m^4_T + m^4_t + m^4_a - 2 m^2_T m^2_t - 2 m^2_T m^2_a- 2 m^2_t m^2_a } \\
&\approx& \frac{1}{2m_T} \sqrt{m^4_T + m^4_t - 2 m^2_T m^2_t }\,.
\end{eqnarray}
If the decay direction of $a$ is perpendicular to its propagation direction, then the angular distance between the decay products of $a$ in the lab frame is:
\begin{eqnarray}
\Delta \theta = 2 \arctan \left( \frac{m_a}{p} \right) \approx \frac{4 m_a m_T}{\sqrt{m^4_T + m^4_t - 2 m^2_T m^2_t}}\,.
\end{eqnarray}
If $\Delta \theta$ is small enough, then most of the objects resulting  from the $a$ decay are clustered in a single jet.
The jet clustering distance parameter $R$ used in Ref~\cite{Sirunyan:2017yta} is $0.4$. 
Here, therefore, we assume that the boosted pseudo-scalar $a$ with $\Delta \theta < 0.2$ is tagged as a single jet. 

\begin{figure}[htbp]
	\begin{center}
		\includegraphics[width=0.50\textwidth]{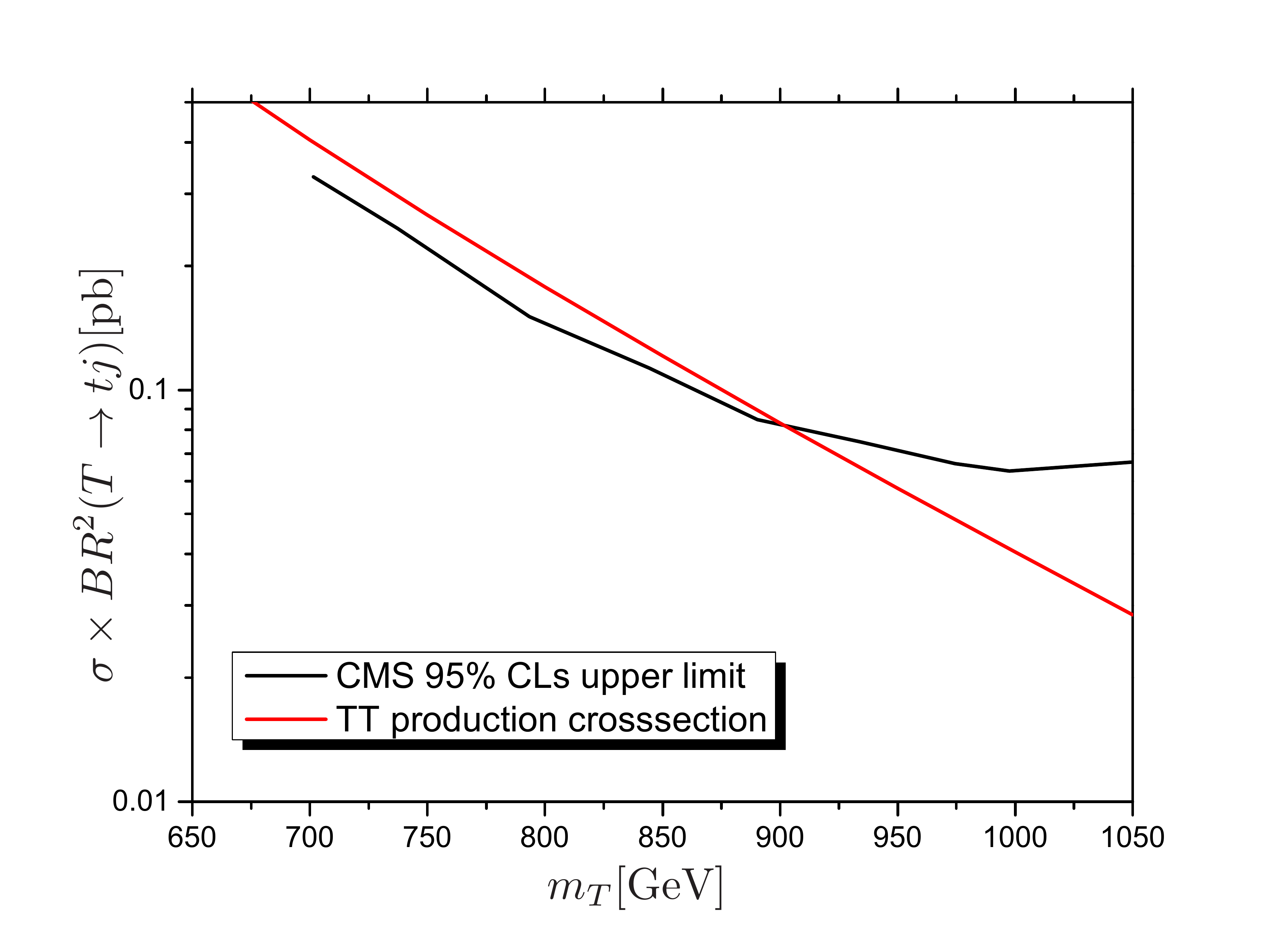}
		\caption{Black line is the observed 95\% $CL_s$ upper limit for the $T$ pair production cross-section times the square of $\mbox{BR}(T\to ta)$ as a function of $m_T$. The red line is the corresponding pair production cross-section at $\sqrt{s} =13$~TeV LHC. }
		\label{ExcitedT}
	\end{center}
\end{figure}

We further assume that the signal efficiency for the $T$ and excited top quark are the same, provided their masses are equal.
Thus, the cross section upper limit given in Ref.~\cite{Sirunyan:2017yta} can be used to directly constrain the exotic $T$ decays. In Fig.~\ref{ExcitedT}  we show the 95\% $CL_s$ cross-section upper limit as function of the $T$ mass compared to the production one: $T$ masses lighter than $910$~GeV are excluded, provided the branching ratio for $T \to t a$ is $100\%$ and $a$ is tagged as a single jet.

%%%%%%%%%%%%%%%%%%%%%%%%%%%%%%%%%%%%%%%%%%%%%%%%%%%%%%%%%%%%%%%%%%%
%%%%%%%%%%%%%%%%%%%%%%%%%%%%%% FIGURE  %%%%%%%%%%%%%%%%%%%%%%%%%%%%%%%%
%%%%%%%%%%%%%%%%%%%%%%%%%%%%%%%%%%%%%%%%%%%%%%%%%%%%%%%%%%%%%%%%%%%
%%%%%%%%%%%%%%%%%%%%%%%%%%%%%%%%%%%%%%%%%%%%%%%%%%%%%%%%%%%%%%%%%%%
%%%%%%%%%%%%%%%%%%%%%%%%%%%%%% FIGURE  %%%%%%%%%%%%%%%%%%%%%%%%%%%%%%%%
%%%%%%%%%%%%%%%%%%%%%%%%%%%%%%%%%%%%%%%%%%%%%%%%%%%%%%%%%%%%%%%%%%%

%\newpage

\subsection{ATLAS $8$~TeV RPV gluino pair search}
A detailed cut flow table is not provided in Ref.~\cite{Aad:2015lea}. We can thus only simulate and compare signal event numbers  after applying the full set of cuts. Table~\ref{Validation1} shows the signal event numbers we obtain in our recast as compared to those provided in Ref.~\cite{Aad:2015lea} (in brackets).

\begin{table}[htbp]
	\begin{center}
		\begin{tabular}{|l|c|c|c|}
			\hline  
			\diagbox{Sample}{SR} & SR1($M^{\Sigma}_{J}$) & \tabincell{c}{$n_{jet}\geqslant7$, \\ $p_T\geqslant120$GeV, \\ $n_b\geqslant1$} & \tabincell{c}{$n_{jet}\geqslant7$, \\ $p_T\geqslant80$GeV, \\ $n_b\geqslant2$} \\
			\hline 
			\tabincell{c}{ $m_{\tilde{g}}=600$GeV,\\$m_{\tilde{\chi}^0_1}=50$GeV} & 85.8(70) & 200.9(180)  & - \\
			\hline 
				\tabincell{c}{ $m_{\tilde{g}}=1000$GeV,\\$m_{\tilde{\chi}^0_1}=600$GeV}&  50.0(55)  & 113.6(101)  & - \\
			\hline 
				\tabincell{c}{ $m_{\tilde{g}}=400$GeV,\\$m_{\tilde{\chi}^0_1}=50$GeV} &  -  & -  & 2135 (1900) \\
			\hline 
			\tabincell{c}{ $m_{\tilde{g}}=500$GeV, \\ BR(t)=0, BR(b)=1 } &  -  & -  & 2038 (1900) \\
			\hline 
			\tabincell{c}{ $m_{\tilde{g}}=500$GeV, \\ BR(t)=1, BR(b)=1  } &  -  & -  & 3159 (3600) \\
			\hline 
			\tabincell{c}{ $m_{\tilde{g}}=600$GeV, \\ BR(t)=1, BR(b)=1  } &  -  & -  & 2131 (2300) \\
			\hline 
		\end{tabular} 
	\caption{Validation of our cut-flow reproduction of Ref.~\cite{Aad:2015lea}. We provide event numbers after the full cut-flow for various benchmark models compared to the ATLAS ones in brackets.} 	\label{Validation1}
	\end{center}

\end{table}

\subsection{ATLAS $13$~TeV RPV gluino pair search}\label{app:RPV13}

\begin{figure}[htb]
	\begin{center}
		\includegraphics[width=0.50\textwidth]{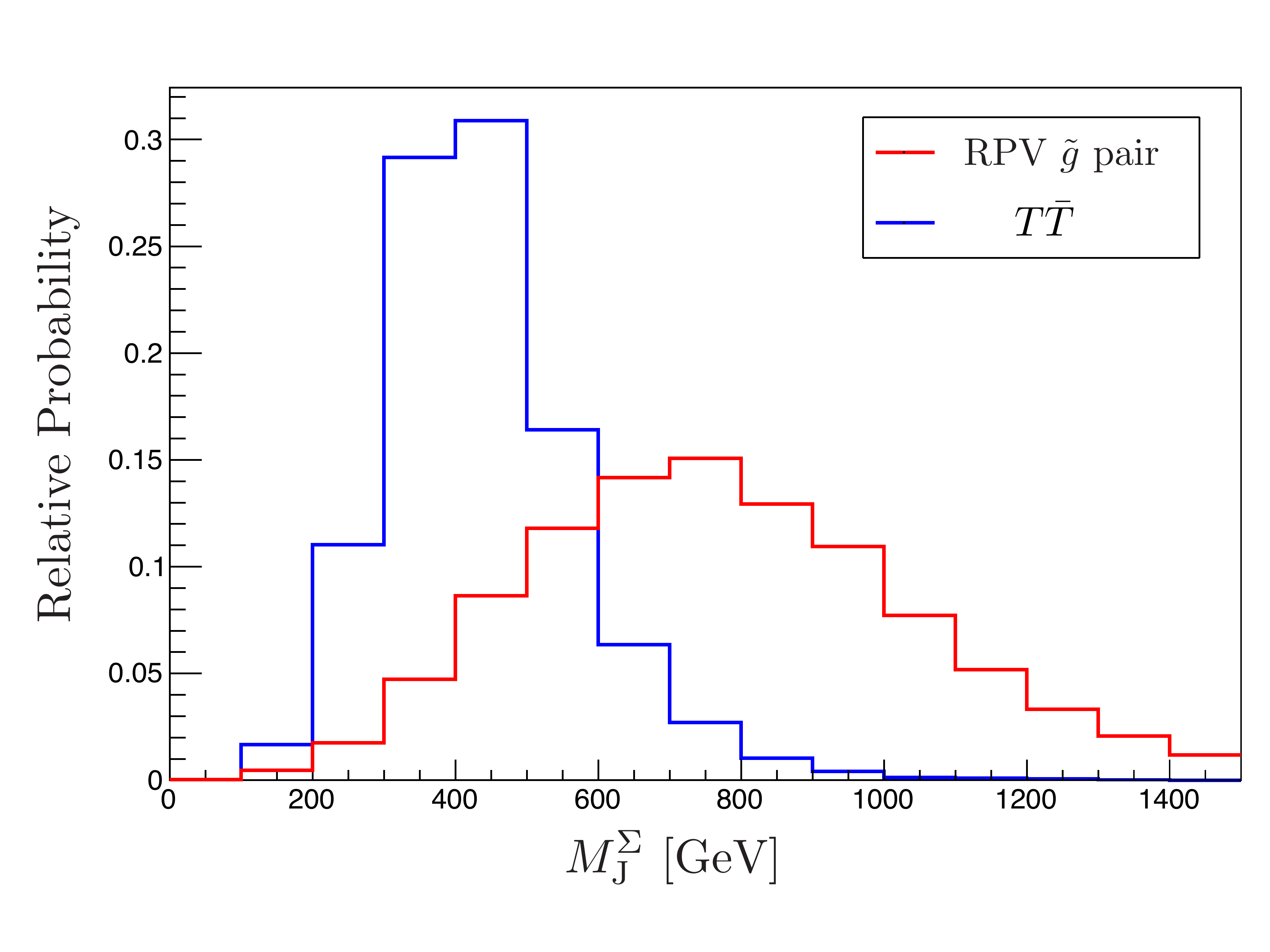}
		\caption{Summed jet mass $M^{\Sigma}_\text{J}$ distribution of a RPV gluino benchmark pair and of $T$ pair. For the gluino signal we use the same benchmark used in Refs~\cite{Aaboud:2018lpl} and ~\cite{Aaboud:2018xuw}:  $m_{\tilde{g}}=1800$~GeV and $m_{\tilde{\chi}^0_1}=1050$~GeV. For  our signal we set $m_{T}=1000$~GeV and $BR(T\to th) = 100\%$. }
		\label{Jetmass}
	\end{center}
\end{figure}

Ref.~\cite{Aaboud:2018lpl} provides a detailed cut flow. Table~\ref{tab:validation2} shows the event numbers after each cut as reported in Ref.~\cite{Aaboud:2018lpl} (in brackets) as well as event numbers we obtain in our recast, showing good agreement. For the signal considered in this article, however, the hard cut on summed jet mass  $M^{\Sigma}_\text{J}$ rejects most signal events.
To illustrate this, we show the $M^{\Sigma}_\text{J}$ distribution of an RPV gluino pair with  $m_{\tilde{g}}=1800$~GeV and $m_{\tilde{\chi}^0_1}=1050$~GeV (a target benchmark of Ref.~\cite{Aaboud:2018lpl}) in comparison to a $T\bar{T}$ pair signal with $m_T=1000$~GeV (a typical mass considered in this article) in Fig. \ref{Jetmass}.
 
\begin{table}[htb]
	\begin{center}
		\begin{tabular}{|l|c|c|cl}
			\hline  
			\diagbox{Cut}{Sample} &  $m_{\tilde{g}}=1800$GeV  & \tabincell{c}{$m_{\tilde{g}}=1800$GeV, \\ $m_{\tilde{\chi}^0_1}=1050$GeV}  \\
			\hline 
			Trigger   & 99.7 (99.7)  & 99.7 (99.7) \\
			\hline 
			$ n_\text{jet} \geqslant $ 4   & 76.6 (74.1)  & 87.8 (88.4) \\
			\hline 
			$|\Delta \eta_\text{12} | > $ 1.4  & 67.9 (74.1) & 79.6 (88.4) \\
			\hline 
			$ n_\text{jet} \geqslant $ 4, $p_\text{T,j1} > $ 400GeV  & 67.9 (74)  & 79.4 (88.4) \\
			\hline 
			\tabincell{c}{$ n_\text{jet} \geqslant $ 4, $p_\text{T,j1} > $ 400GeV,  \\ $M^{\Sigma}_\text{J} > $  1.0 TeV}  & 5.6 (7.38) & 17.0 (25.4) \\
			\hline 
			\tabincell{c}{$ n_\text{jet} \geqslant $ 4, $p_\text{T,j1} > $ 400GeV, \\ $N_\text{b-jet} > $ 0}   & 52.3 (52.9) & 61.1 (69.6) \\
			\hline 
			\tabincell{c}{$ n_\text{jet} \geqslant $ 4, $p_\text{T,j1} > $ 400GeV, \\ $N_\text{b-jet} > $ 0, $M^{\Sigma}_\text{J} > $  1.0 TeV}  & 4.3 (5.3)  & 13.1 (19.9) \\
			\hline 
			$ n_\text{jet} \geqslant $ 5    & 30.5 (31.3)  & 48.1 (54.8) \\
			\hline 
			$ n_\text{jet} \geqslant $ 5, $M^{\Sigma}_\text{J} > $  0.8 TeV   & 3.9 (5.4)  & 17.7 (26.9) \\
			\hline 
			$ n_\text{jet} \geqslant $ 5,  $N_\text{b-jet} > $ 0  & 23.5 (22.6)  & 37.0 (43.5) \\
			\hline
			\tabincell{c}{$ n_\text{jet} \geqslant $ 5,  $N_\text{b-jet} > $ 0, \\ $M^{\Sigma}_\text{J} > $  0.8 TeV }   & 3.0 (3.9)  & 13.6 (21.4) \\
			\hline  
		\end{tabular} 
	\caption{Validation of our cut-flow reproduction of Ref.~\cite{Aaboud:2018lpl}. We provide event numbers after each cut for two benchmark models compared to the ATLAS ones in brackets. }  \label{tab:validation2}
	\end{center}

\end{table}

%%%%%%%%%%%%%%%%%%%%%%%%%%%%%%%%%%%%%%%%%%%%%%%%%%%%%%%%%%%%%%%%%%%
%%%%%%%%%%%%%%%%%%%%%%%%%%%%%% FIGURE  %%%%%%%%%%%%%%%%%%%%%%%%%%%%%%%%
%%%%%%%%%%%%%%%%%%%%%%%%%%%%%%%%%%%%%%%%%%%%%%%%%%%%%%%%%%%%%%%%%%%
%%%%%%%%%%%%%%%%%%%%%%%%%%%%%%%%%%%%%%%%%%%%%%%%%%%%%%%%%%%%%%%%%%%
%%%%%%%%%%%%%%%%%%%%%%%%%%%%%% FIGURE  %%%%%%%%%%%%%%%%%%%%%%%%%%%%%%%%
%%%%%%%%%%%%%%%%%%%%%%%%%%%%%%%%%%%%%%%%%%%%%%%%%%%%%%%%%%%%%%%%%%%

\subsection{ATLAS $13$~TeV up-type vector-like quark search} \label{app:VLQ}

A detailed cut flow table is not provided in Ref.~\cite{Aaboud:2018xuw}. We can thus only simulate and compare signal event numbers  after applying the full set of cuts. We focus on the 1-lepton signal regions, which are most sensitive and for which signal efficiency and background information is provided.
Table~\ref{Validation3} shows the signal event numbers we obtain in our recast as compared to those provided in Ref.~\cite{Aaboud:2018xuw} (in brackets). 

The expected signal event number is sensitive to the multi b-jet tagging. The discrepancy between our result and Ref.~\cite{Aaboud:2018xuw} exceeds $20\%$ in some cases, and our expected signal numbers are always smaller. Also, we do not combine bounds from different search regions and only report the largest bound from a single search region. Thus our recast  underestimates the bound. For $T\bar{T}$ pair production with $100\%$ branching ratio $T\rightarrow th$, we obtain a recast expected bound of $m_T > 1050$~GeV, while \cite{Aaboud:2018xuw} reports an expected bound of  $m_T > 1340$~GeV when combining all 0- and 1-lepton search regions.

\begin{figure}[htb]
	\begin{center}
		\includegraphics[width=0.50\textwidth]{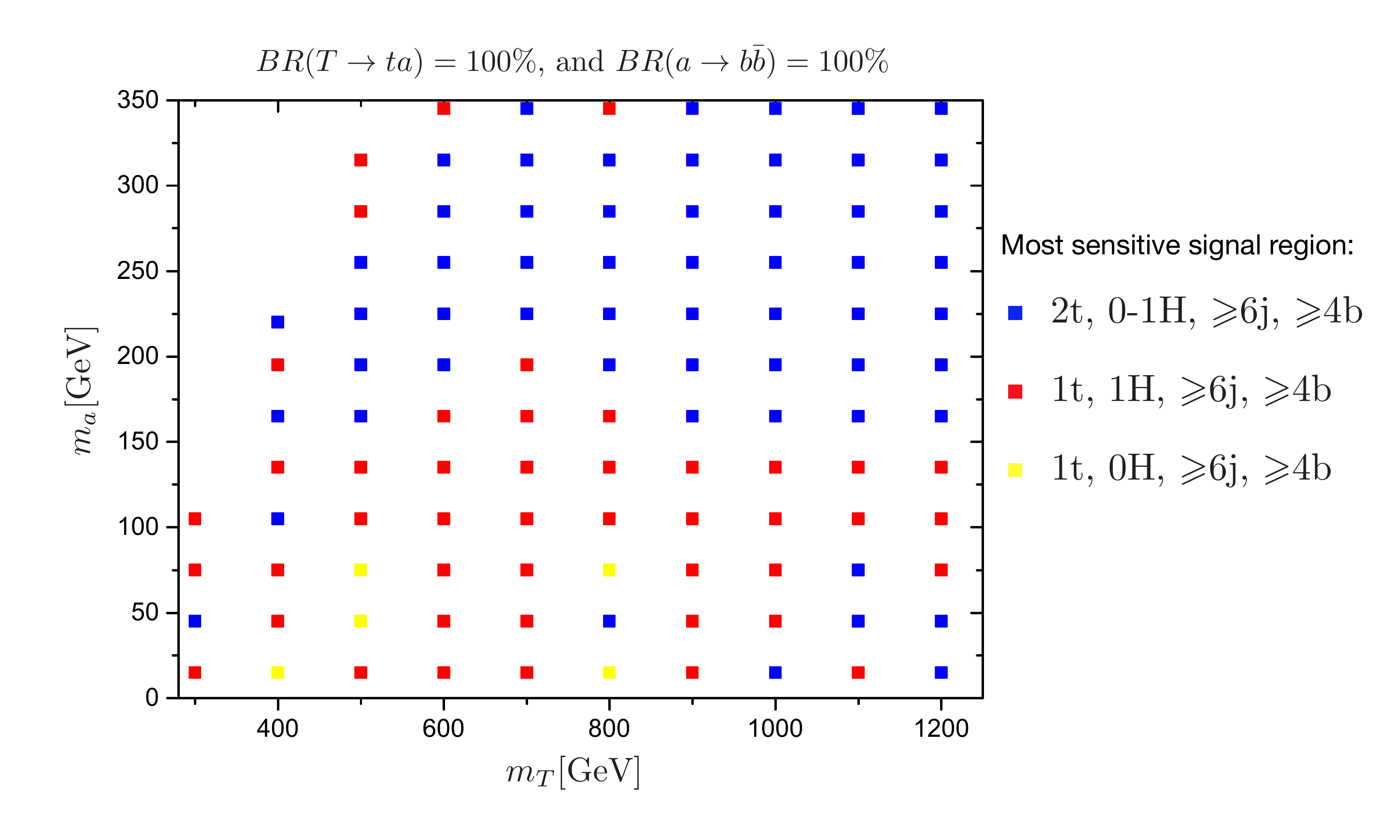}
		\caption{Most sensitive signal region for different $m_T$ and $m_a$. Here we assume $\mbox{BR}(T\to ta) = 100\%$ and $\mbox{BR}(a\to b\bar{b}) = 100\%$.}
		\label{Most}
	\end{center}
\end{figure}

Fig. \ref{Most} shows the most sensitive signal region for different $m_T$ and $m_a$.

\begin{table}[htb]
	\begin{center}
		\begin{tabular}{|c|c|c|c|c|}
			\hline
			\multicolumn{5}{|c|}{1-lepton channel, $m_T = 1$~TeV,  $\mbox{BR}(T \to th) = 100\%$} \\
			\hline  
			\tabincell{c}{ $\geqslant$2t, 0-1H, \\ $\geqslant$6j, 3b } &  \tabincell{c}{ 1t, 0H, \\ $\geqslant$6j, $\geqslant$4b }&  \tabincell{c}{ 1t, 1H, \\ $\geqslant$6j, $\geqslant$4b }&  \tabincell{c}{ $\geqslant$2t, 0-1H, \\ $\geqslant$6j, $\geqslant$4b }&  \tabincell{c}{ $\geqslant$0t, $\geqslant$2H, \\ $\geqslant$6j, $\geqslant$4b }  \\
			\hline 
			   \tabincell{c}{ 18.3\\(19.6) }&  \tabincell{c}{ 15.6\\(21.5) }&  \tabincell{c}{ 15.3\\(24.3) }&  \tabincell{c}{ 17.8\\(23.9) }&  \tabincell{c}{ 9.7\\(14.6) } \\
			\hline 
		\end{tabular} \caption{Validation of our cut-flow reproduction of Ref.~\cite{Aaboud:2018xuw}. We provide event numbers after the full cut-flow for various benchmark models compared to the ATLAS ones in brackets. }	\label{Validation3}
	\end{center}
\end{table}

%\begin{table*}[htb]
%	\begin{center}
%		\begin{tabular}{|c|c|c|c|c|c|}
%			\hline  
%			1-lepton channel &  \tabincell{c}{ $\geqslant$2t, 0-1H, \\ $\geqslant$6j, 3b } &  \tabincell{c}{ 1t, 0H, \\ $\geqslant$6j, $\geqslant$4b }&  \tabincell{c}{ 1t, 1H, \\ $\geqslant$6j, $\geqslant$4b }&  \tabincell{c}{ $\geqslant$2t, 0-1H, \\ $\geqslant$6j, $\geqslant$4b }&  \tabincell{c}{ $\geqslant$0t, $\geqslant$2H, \\ $\geqslant$6j, $\geqslant$4b }  \\
%			\hline 
%			\tabincell{c}{$m_T =$ 1 TeV, \\ $BR(T \to Ht) = 1.0$  }   &  \tabincell{c}{ 18.3\\(19.6) }&  \tabincell{c}{ 15.6\\(21.5) }&  \tabincell{c}{ 15.3\\(24.3) }&  \tabincell{c}{ 17.8\\(23.9) }&  \tabincell{c}{ 9.7\\(14.6) } \\
%			\hline 
%		\end{tabular} \caption{Validation of our cut-flow reproduction of Ref.~\cite{Aaboud:2018xuw}. We provide event numbers after the full cut-flow for various benchmark models compared to the ATLAS ones in brackets. }	\label{Validation3}
%	\end{center}
%\end{table*}

%%%%%%%%%%%%%%%%%%%%%%%%%%%%%%%%%%%%%%%%%%%%%%%%%%%%%%%%%%%%%%%%%%%
%%%%%%%%%%%%%%%%%%%%%%%%%%%%%% FIGURE  %%%%%%%%%%%%%%%%%%%%%%%%%%%%%%%%
%%%%%%%%%%%%%%%%%%%%%%%%%%%%%%%%%%%%%%%%%%%%%%%%%%%%%%%%%%%%%%%%%%%
%%%%%%%%%%%%%%%%%%%%%%%%%%%%%%%%%%%%%%%%%%%%%%%%%%%%%%%%%%%%%%%%%%%
%%%%%%%%%%%%%%%%%%%%%%%%%%%%%% FIGURE  %%%%%%%%%%%%%%%%%%%%%%%%%%%%%%%%
%%%%%%%%%%%%%%%%%%%%%%%%%%%%%%%%%%%%%%%%%%%%%%%%%%%%%%%%%%%%%%%%%%%

\newpage

\bibliographystyle{utphys}
\bibliography{eVLQbds}

\end{document}